\documentclass[namedreferences]{solarphysics}
\usepackage[optionalrh]{spr-sola-addons} 
\usepackage{graphicx}        
\usepackage{color}           
\usepackage{url}             

\newcommand{\ie}{{\it i.e.}}
\newcommand{\eg}{{\it e.g.}}

\begin{document}

\begin{article}

\begin{opening}

\title{Non-Thermal ``Burst-on-Tail'' of Long-Duration Solar Event on 26 October 2003\\ {\it Solar Physics}}

\author{I.~\surname{Zimovets}$^{1}$\sep
        A.~\surname{Struminsky}$^{1}$\sep      
       }
\runningauthor{Zimovets \& Struminsky}
\runningtitle{Burst-on-Tail Event on 26 October 2003}

   \institute{$^{1}$ Space Research Institute (IKI), Russian Academy of Sciences, 117997, 84/32 Profsoyuznaya Str, Moscow, Russia
                     email: \url{ivanzim@iki.rssi.ru}, email: \url{astrum@iki.rssi.ru}\\ 
                           }

\begin{abstract}
Observations of a rare long-duration solar event of GOES class X1.2 from 26 October 2003 are presented. This event showed a pronounced burst of hard X-ray and microwave emission, which was extremely delayed ($>$60~min) with respect to the main impulsive phase and did not have any significant response visible in soft X-ray emission. We refer to this phenomenon as a ``burst-on-tail''. Based on TRACE observations of the growing flare arcade and some simplified estimation, we explain why a reaction of active region plasma to accelerated electrons may change drastically over time. We suggest that, during the ``burst-on-tail'', non-thermal electrons were injected into magnetic loops of larger spatial scale than during the impulsive phase bursts, thus resulting in much smaller values of plasma temperature and emission measure in their coronal volume, and hence little soft X-ray flux. The nature of the long gap between the main impulsive phase and the ``burst-on-tail'' is, however, still an open question.
\end{abstract}
\end{opening}

\section{Introduction}
     \label{S-Introduction} 

Physical links between thermal and non-thermal electromagnetic emission in solar flares are not yet fully understood. There is an important observational fact, called the ``Neupert effect'' by \inlinecite{Hudson91}, which phenomenologically relates thermal soft X-ray (SXR) emission with non-thermal hard X-ray (HXR) and microwave emission of solar (and stellar) flares. According to this effect, often the time integral of HXR and microwave light curves closely resembles the time profile of SXR emission during a flare's rising phase (\eg, \opencite{Neupert68}; \opencite{Kahler70}; \opencite{Dennis93}; \opencite{Hawley95}; \opencite{Gudel96}; \opencite{Veronig05}). The ``Neupert effect'' supports a scenario in which SXR emitting plasma consists mainly of evaporated chromospheric material heated to several MK by the same population of accelerated and precipitated electrons that radiate non-thermal emission in the solar atmosphere.

However, in some flares the ``Neupert effect'' is not observed (\eg, \opencite{Feldman90}; \opencite{Veronig02a}). This is especially pronounced in flares where there is brightly marked pre-impulsive phase heating observed in the SXR range which lacks a counterpart in HXR emission \cite{Veronig02b}, and also in large non-impulsive X- and M-class flares with gradually varying HXR emission. These latter events are referred to as gradual-hard flares or long-duration events (LDEs) (\eg, \opencite{Feldman90}; \opencite{Dennis93}).

Recently \inlinecite{Warmuth09} have reported an intriguing X1.3-class two-ribbon flare with a couple of intense bursts of non-thermal HXR emission, which were separated by about 2 min from a group of impulsive phase bursts. The Neupert effect was valid only for the impulsive phase bursts, but not for the delayed pair. These delayed bursts were even more intense than the impulsive phase bursts in the energy range above $h\nu \approx50$~keV, but they were observed during the decay phase of SXR emission. Spectral analysis of HXR emission observed by the \textit{Reuven Ramaty High-Energy Solar Spectroscopic Imager} (RHESSI; \opencite{Lin02}) allowed \inlinecite{Warmuth09} to suggest the presence of an anomalously high low-energy cutoff of non-thermal electrons ($E_{\rm{c}}\approx 100$~keV) during the delayed bursts, whereas in the impulsive phase bursts the low-energy cutoff was estimated as only $E_{\rm{c}}\approx 30-40$~keV. Thus, it was concluded by \inlinecite{Warmuth09} that the lack of appreciable plasma heating during the delayed bursts could be due to relatively small kinetic power of non-thermal electrons in comparison with that of the impulsive phase bursts. However, the existence of such a huge low-energy cutoff ($E_{\rm{c}}\approx100$~kev) in a spectrum of accelerated electrons in solar flares is an open question (\eg, \opencite{Holman11}; \opencite{Kontar11}).             
  
The goal of this paper is to present another example of electron acceleration without visible response in SXR emission and propose its possible interpretation. A highly pronounced burst of HXR and microwave emission was observed in the decay phase of the X1.2-class LDE on 26 October 2003 with a delay of more than 60 min relative to the main impulsive phase (Figure~\ref{fig1}). Hereafter we will refer to this burst as the ``burst-on-tail''. To our best knowledge, the LDE of 26 October 2003 is the first event of its kind ever reported in the literature. It might represent a new type of flare phenomenon. 

\section{Observations}
 \label{S-Obs}

\subsection{Instruments and Data}
 \label{SS-Instr}

\subsubsection{SXR and HXR Emission}
 \label{XRE}

The principal instrument used in this work is the Anti-Coincidence System (ACS) of the spectrometer onboard INTEGRAL (SPI). The ACS is made of a set of BGO crystals viewed by photomultipliers. The total mass of these crystals is 512~kg with an effective area of 0.3~m$^2$. The ACS is a very sensitive instrument to both primary and secondary gamma-rays with energies $h\nu \gtrsim 100$~keV. The time resolution of the ACS is 50~ms, but here we use 1-s count rate data that is sufficient for our purpose. A more detailed description of the ACS and SPI can be found, \eg{}, in \inlinecite{Vedrenne03}. The ACS may continuously monitor solar high energy activity for a period of more than two days, due to its eccentric orbit with an apogee of $\approx150\,000$~km and the 72-h orbiting period of the INTEGRAL spacecraft. Thus, the ACS is a unique instrument for studying long-duration solar events, despite its inability to perform spectrally and spatially resolved observations. The ACS is not a solar-dedicated instrument, but it was already succesfully used for solar studies (\eg, \opencite{Gros04}; \opencite{Kiener06}; \opencite{Su06}; \opencite{Struminsky10}). 

There are no usable RHESSI observational data for almost the entire time interval of interest to us, \ie{} from about 06:00 UT until 08:00 UT of 26 October 2003 (see Figure~\ref{fig1}). During the ``burst-on-tail'' (about 07:25--07:40~UT), which is the main object of our study, RHESSI (according to Murphy's law) was in the South Atlantic Anomaly. Therefore, a spectral analysis of the HXR emission during the event is not possible. 

We also use 3-s data of the two standard GOES SXR channels - 1--8 \r{A} and 0.5--4 \r{A}  - to calculate time profiles of plasma emission measure and temperature in the flare region \cite{Thomas85}. The GOES package of SolarSoftWare is used for these purposes. 

\subsubsection{Radio Emission}
 \label{RE}

To investigate solar microwave emission produced during the event we use time profiles of flux density measured with a time cadence of 0.1 s at six frequencies of 1, 2, 3.75, 9.4, 17, and 35~GHz by the Nobeyama Radio Polarimeters (NoRP; \opencite{Nakajima85}). Unfortunately, the Nobeyama Radioheliograph finished its observations of the Sun at about 06:30~UT, prior to the ``burst-on-tail''.  

In addition to the NoRP datasets, we use time profiles of flux density of solar radio emission at eight discrete standard frequencies of 245, 410, 610, 1415, 2695, 4995, 8800, and 15400~MHz provided by the Learmonth solar radio telescope (RSTN). However, it should be noted here that the Learmonth microwave data were intermittent and anomalously noisy during the ``burst-on-tail'' in comparison with the NoRP data. Therefore, we do not use them for detailed spectral analysis during the ``burst-on-tail''. 

The Learmonth Solar Observatory also provides the radio spectrogram of the event in the frequency range of 25--180~MHz, which is swept every 3 s. This allows us to check metric radio emission of the Sun during the event. 

\subsubsection{UV and EUV Emission}
 \label{OE}
 
We analyze morphology and dynamics of magnetoplasma structures of the flare region in both the chromosphere and in the corona during this event. This is achieved using images obtained by the \textit{Transition Region and Coronal Explorer} (TRACE; \opencite{Handy99}) mainly at 195~\r{A} and 1600~\r{A} with a variable time cadence of several seconds. 

In addition to the UV and EUV observations by TRACE we use observations by Extreme-Ultraviolet Imaging Telescope (EIT; \opencite{Delaboudiniere95}) onboard the \textit{Solar and Heliospheric Observatory} (SOHO) spacecraft. Despite the fact that EIT had a cadence of only 12 min, it has a much larger field of view ($45 \times 45$~arcmin) than TRACE ($6.4 \times 6.4$~arcmin) and thus covers the entire visible solar surface. This allows us to check that both the impulsive phase bursts and the ``burst-on-tail'' took place in the same active region. 

Magnetic topology of the flare region was clarified by the line-of-sight photospheric magnetograms produced by the Michelson Doppler Imager (MDI; \opencite{Scherrer95}) onboard SOHO.

\begin{figure}
\centerline{\includegraphics[width=0.98\textwidth, bb=2 330 513 818, clip=]{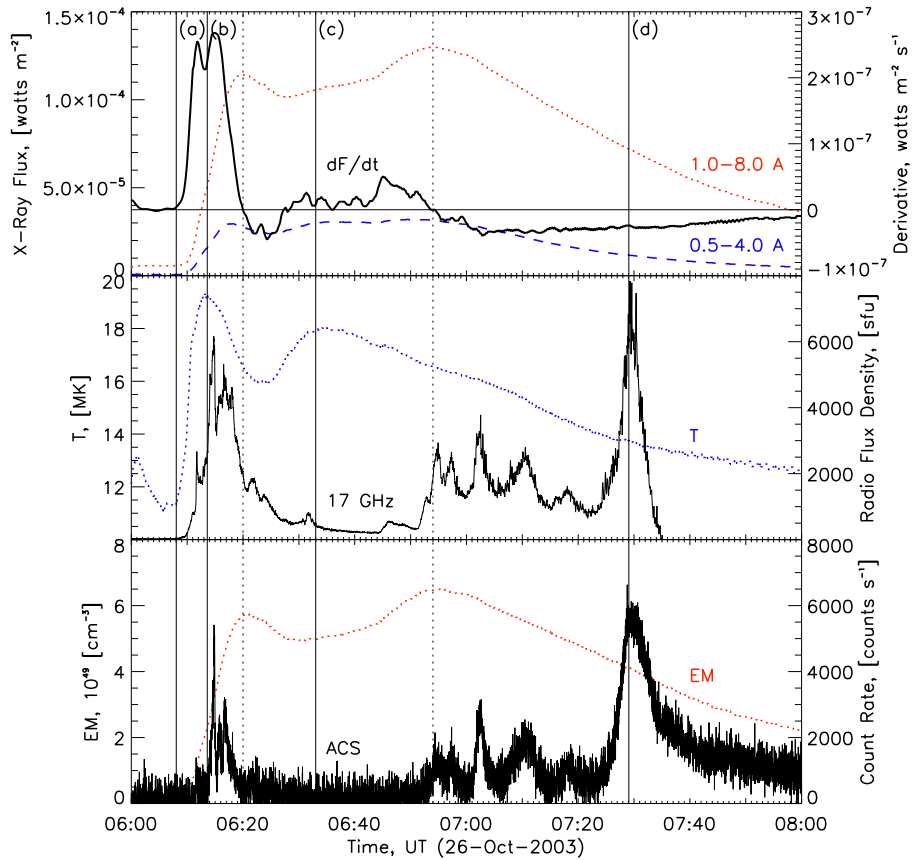}
            }
\caption{Time profiles of X-ray and microwave emission as well as of some estimated parameters of the radiated plasma for the solar flare of 26 October 2003. Upper panel: Fluxes of X-ray emission (left axis) detected by GOES in two spectral channels 1.0-8.0 \r{A} (red dotted line) and 0.5-4.0 \r{A} (blue dashed line). The black thick solid line indicates the smoothed time derivative of X-ray flux in the 1.0-8.0 \r{A} channel (right axis), while the horizontal thin solid line shows the zero level of this time derivative. Middle panel: Temperature of the flare plasma estimated from the GOES data (left axis, blue dotted line) and flux density of solar radio emission detected by NoRP at 17~GHz (right axis, black solid line). Bottom Panel: Emission measure of the flare plasma estimated from the GOES data (left axis, red dotted line) and count rates from the ACS (right axis, black solid line). Four vertical solid lines with notations (a), (b), (c), and (d) mark the moments for which the images presented in Figure~\ref{fig2} were made. The vertical dotted lines indicate the moments of the first (06:20~UT) and the second (06:54~UT) soft X-ray maxima. 
}
\label{fig1}
\end{figure}

\subsection{Dynamics of the Flare Electromagnetic Emission}

The X1.2 solar flare of 26 October 2003 originated from NOAA active region 10486. Its center was at a point with heliographic coordinates S14E41. Owing to this east location solar cosmic rays, which could be accelerated during the event, did not reach the Earth and did not interfere with the detection of solar HXR emission by the ACS. 

The event started at about 05:57~UT and peaked at 06:54~UT according to the GOES SXR observations. The decay to the pre-flare level took more than 9 h. A pronounced impulsive phase started at about 06:10~UT and ended at about 06:30 UT (Figure~\ref{fig1}). It was accompanied by almost simultaneous bursts of HXR and microwave emission. A gap in microwave and HXR emission was observed respectively by the NoRP and ACS between about 06:30 and 06:50~UT. A new sequence of intense bursts of HXR and microwave emissions started at 06:50~UT. This sequence of gradual bursts ended in the strong ``burst-on-tail'', which began at 07:20~UT and lasted for more than 20 min. Surprisingly, the ``burst-on-tail'' has the largest HXR and microwave intensities of all the bursts in this event.     

Light curves of SXR emission detected by GOES had two maxima (Figure~\ref{fig1}). The first maximum was at about 06:20~UT. The time derivative of SXR intensity around the first maximum corresponds to the observed intensities of HXR and microwave emission, consistent with the Neupert effect. The second and major SXR maximum at 06:54~UT was just prior to the sequence of delayed (gradual) bursts of HXR and microwave emission. This suggests that softer energy release and plasma heating occurred after the flare impulsive phase but before the delayed bursts. These softer episodes of energy release were highlighted by multiple peaks of the SXR time derivative between about 06:30~UT and 06:50~UT (see Figure~\ref{fig1}), in accordance with the Neupert effect.      
 
However, the Neupert effect was not observed for the delayed sequence of non-thermal bursts, the last and strongest of which was the ``burst-on-tail''. An intriguing fact is that this ``burst-on-tail'' was not associated with a visible increase of SXR fluxes and, therefore, increases of temperature and emission measure (Figure~\ref{fig1}). In reality some weak increase of SXR flux was probably produced during the ``burst-on-tail'', but was hidden by a relatively strong decaying tail after the second SXR maximum.

\begin{figure}
\centerline{\includegraphics[width=0.98\textwidth, bb=75 365 500 700, clip=]{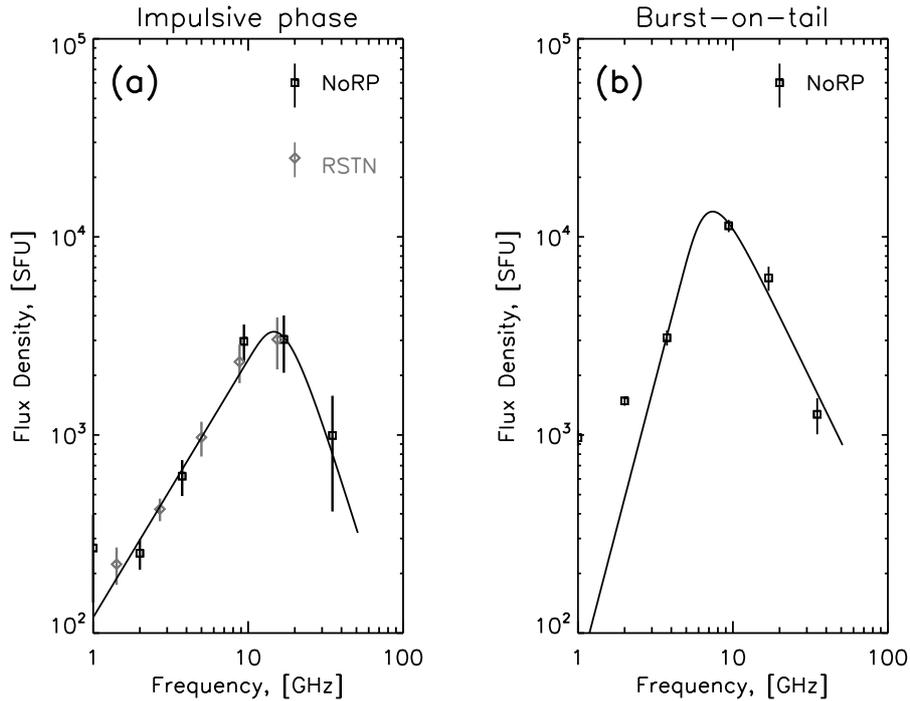}
            }
\caption{Spectra of solar microwave emission in two particular time intervals during the 26 October 2003 event: (a) 06:12:56--06:14:16~UT in the impulsive phase, (b) 07:28:24--07:29:44~UT during the ``burst-on-tail''. The centers of these intervals correspond to two vertical lines in Figure~\ref{fig1} marked by (b) and (d), respectively. Observational data points are plotted by black squares (NoRP) and grey diamonds (RSTN) with vertical line bars marking errors of measurement at the one sigma level. The solid black line represents an approximation of the data points with the function $I\left(\nu \right)=a_{0}\nu^{a_{1}}\left[1-\exp\left(-a_{2}\nu^{-a_{3}}\right)\right]$ . 
}
\label{fig2}
\end{figure}

\subsection{Spectral Analysis of the Flare Microwave Emission}

As was already noted above, HXR spectra for this event were not available. Nevertheless, we can make a rough spectral analysis of the microwave emission using observations by the NoRP and the Learmonth solar radio telescope. As an example, in Figure~\ref{fig2} we plot the microwave spectra for two principal time intervals. Figure~\ref{fig2}(a) represents a spectrum of microwave bursts in the impulsive phase integrated over 06:12:56--06:14:16~UT.  The center of this interval is marked by a vertical line (b) in Figure~\ref{fig1}. Figure~\ref{fig2}(b) shows a microwave spectrum of the ``burst-on-tail'' peak integrated over the time interval 07:28:24--07:29:44~UT. Its center is marked by a vertical line (d) in Figure~\ref{fig1}.

It is clearly seen that both cases of the microwave spectra represented in Figure~\ref{fig2} have a $\Lambda$-shape above a frequency of about 2~GHz. This shape of microwave spectrum is typical for the gyrosynchrotron radiation of a single population of non-thermal electrons in the magnetic field of an active region (\eg, \opencite{Dulk82}; \opencite{Stahli89}). It is well approximated by a function $I\left(\nu \right) = a_{0}\nu^{a_{1}}\left[1-\exp\left(-a_{2}\nu^{-a_{3}}\right)\right]$, where $\nu$ is a frequency, and $a_{0}$, $a_{1}$, $a_{2}$, and $a_{3}$ are some positive parameters, which can change during the course of a flare. This is direct evidence that the microwave emission of the ``burst-on-tail'' was produced by non-thermal electrons. Since the microwave and HXR intensities observed during the ``burst-on-tail'' were well correlated, we can infer that they were generated by the same population of non-thermal electrons.

An upward bend in the microwave spectrum below $\nu \approx 2$~GHz was observed during the ``burst-on-tail'' in addition to its $\Lambda$-shape. Our inspection of the radio spectrograms obtained at the Learmonth Solar Observatory and the reports given in the \textit{Solar Geophysical Data} catalogues shows that a type IV radio burst and multiple decimetric bursts were observed after the impulsive phase. Most probably, the upward bend was due to these bursts.

\begin{figure}
\centerline{\includegraphics[width=0.98\textwidth, bb=60 370 532 832, clip=]{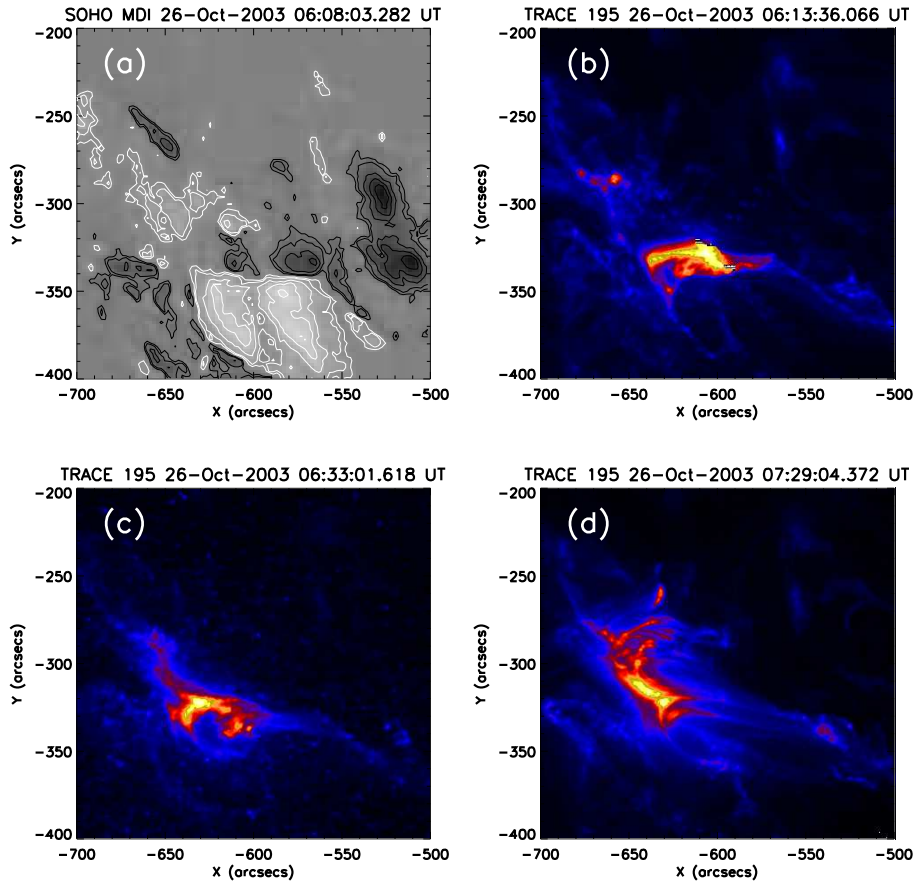}
            }
\caption{Images of NOAA active region 10486 during the 26 October 2003 X1.2 solar flare. (a) Photospheric line-of-sight magnetogram observed by SOHO/MDI. The white and black lines are iso-contours at $\pm 300, 500, 1000, 1500,$ and $2000$~G levels. (b)--(d) Images of the active region made by TRACE at 195~\r{A}. The time of observation of each of these four images are shown in the titles and are marked by vertical dashed lines and also by appropriate letters in parentheses in Figure~\ref{fig1}. 
}
\label{fig3}
\end{figure}

\subsection{Morphology and Dynamics of the Flare Region}

Unfortunately, there were no spatially-resolved observations of the flare region in either the SXR or HXR ranges. However, the flare region was well observed in the UV and EUV ranges by TRACE and SOHO/EIT between 06:00~UT and 08:00~UT. We also checked the entire solar disc and did not find significant flaring activity in other active regions during this time interval. The event was a complex two-ribbon flare, which began with a set (an arcade) of relatively compact flaring loops of lengths $L_{\rm{loop}} \approx \left(3 - 4\right) \times 10^{9}$~cm. In the course of the event new magnetic loops gradually became involved in the flaring process. The flare arcade grew both along the magnetic polarity inversion line, and across it. Hence the spatial scale of magnetic loops of the flare arcade became larger and larger with time. Their apparent lengths reached $L_{\rm{loop}} \approx \left(7 - 14\right) \times 10^{9}$~cm during the ``burst-on-tail''. Figures~\ref{fig3}(b), \ref{fig3}(c), and \ref{fig3}(d) show this process dynamically. These three panels correspond to the times marked by the same letters in parentheses and by vertical lines in Figure~\ref{fig1}.    

Figure~\ref{fig3}(a) represents the line-of-sight photospheric magnetogram of the flare region observed by MDI just prior to the flare impulsive phase. It is clearly seen that the active region had a complex bipolar structure, which was formed by two sunspots of positive polarity in the south and several sunspots of negative polarity in the northwest. Footpoints of the flare magnetic loops were situated in this sheared complex bipolar structure.

For further discussion it is also important that this was an eruptive event. It was accompanied by quite a fast ($v>10^{8}$~cm~s$^{-1}$) coronal mass ejection (CME) observed by the LASCO/C2 coronagraph onboard SOHO for the first time at 06:54:30~UT, just at the major soft X-ray maximum of the flare\footnote{http://cdaw.gsfc.nasa.gov/CME\_list/UNIVERSAL/2003\_10/univ2003\_10.html}. 

\section{Discussion}
 \label{S-Disc}

The implemented analysis of the long-duration solar event of 26 October 2003 provides observational evidence of long-lasting ($\approx1.5$~h) flaring activity with multiple acceleration episodes of electrons. It is clearly demonstrated that a strong burst of non-thermal HXR and microwave emission, which had even greater intensity than the flare impulsive phase, appeared during the decay phase of the flare, and originated from the same NOAA active region as the impulsive emission. This non-thermal ``burst-on-tail'' had two pronounced features, which we would like to emphasize here: 
\begin{enumerate}
\item{The ``burst-on-tail'', despite being more intense than the flare impulsive emission, had no significant counterpart in SXR emission.}  
\item{The spatial scale of the flaring loops during the ``burst-on-tail'' was several times larger than during the impulsive phase bursts.}
\end{enumerate}

\subsection{Weak Soft X-ray Emission during the ``Burst-on-Tail''}

There are at least two possible interpretations of relatively weak soft X-ray emission during the ``burst-on-tail'' with respect to the flare impulsive phase. As discussed by \inlinecite{Warmuth09} for the 19 January 2005 event, a hypothetical abrupt increase of the low-energy cutoff of nonthermal electrons during the ``delay'' episode of the flare energy release is possible. However, we cannot easily test this interpretation for the 26 October 2003 event, since the RHESSI data needed for a detailed spectral analysis of HXR emission are not available. Hence we are not able to estimate whether the low-energy cutoff could have changed during the flare. Moreover an extraction of the low-energy cutoff of non-thermal electrons from even a high quality HXR spectrum remains questionable and often does not have a unique solution (\eg, \opencite{Holman11}; \opencite{Kontar11}; and references therein).    

An ideologically similar explanation is based on a possible hardening of the non-thermal electron spectrum over the course of the event, which is a common feature of large solar flares (\eg, \opencite{Frost71}; \opencite{Kiplinger95}; \opencite{Grigis08}; \opencite{Saldanha08}). It is obvious that if the amount of electrons accelerated in two episodes of energy release is the same, but their spectrum is different, then the number of electrons with higher energies (say, with $E\gtrsim50$~keV) is larger in the episode with the harder spectrum, while there are fewer electrons with lower energies. Thus, fluxes of HXR and microwave radiation would be larger, and fluxes of SXR emission would be smaller for a harder electron spectrum. Similarly, this hypothesis cannot be tested here due to the lack of HXR observations. 

We propose an alternative interpretation. The main idea is that a population of non-thermal electrons injected into a large flare loop should lead to less SXR response of its plasma in comparison with the injection into a smaller loop. In our case the flare energy release during the non-thermal ``burst-on-tail'' corresponded to a larger loop scale than during the impulsive phase bursts. The growth of flare loops is a well-known property of solar eruptive events, which is explained naturally by the ``standard'' model (\eg, \opencite{Priest02}; \opencite{Hudson11}; \opencite{Shibata11}; and references therein).  

Based on this idea, we can conduct further qualitative analysis. The goal is to estimate a plasma response of two flaring loops having different lengths to the same population of non-thermal electrons. For simplicity, we will assume that the observed flare loop plasma, which emits in the SXR range, is heated by nonthermal electrons only. All other possible types of heating are neglected. The population of nonthermal electrons has a power-law energy distribution function $F\left(E\right) = A E^{-\delta}$ with the low-energy cutoff $E_{\rm{c}}$.
 
Firstly, let us estimate a fraction $\alpha \approx I_{1} / I_{0}$ of the total energy \\ $I_{0} = A \int^{+\infty}_{E_{\rm{c}}} E^{1-\delta} dE = A E_{\rm{c}}^{2-\delta} / \left(\delta-2\right)$, which non-thermal electrons lose in the coronal part of the flare loops due to Coulomb collisions with plasma during their precipitation to the chromospheric footpoints. After propagation through the column density $N$~cm$^{-2}$ an electron with initial energy $E$ has energy 
\begin{equation}
 E_{1} \simeq E \left(1-2KNE^{-2} \right)^{1/2}, 
\label{eq1}
\end{equation}
where $K=2 \pi e^{4} \lambda_{\rm{ee}}$, $e$ is the electron charge, and $\lambda_{\rm{ee}}$ is the electron-electron collision Coulomb logarithm (\eg, \opencite{Jackson62}; \opencite{Brown72}). For typical physical conditions of flare regions $K\approx8\times10^{-36}$~erg$^{2}$~cm$^{2}$. For simplicity we assume the plasma density $n$ is constant along the coronal part of the flare loop. Let us compare possible values of $n$ with a quantity $E^{2}/KL$, where $L$ is the length traversed by an electron. If all nonthermal electrons have energy greater than $E_{\rm{c}}\approx20$~keV, which seems a reasonable value of the low-energy cutoff (\eg, \opencite{Holman11}; \opencite{Kontar11}), and $L \approx L_{\rm{loop}}/2 \approx \left(2 - 7\right)\times10^{9}$~cm, where $L_{\rm{loop}}$ is the characteristic loop length observed by TRACE, then we find $E_{\rm{c}}^{2}/KL\approx\left(2 - 6 \right)\times10^{10}$~cm$^{-3}$. Unfortunately, we cannot estimate $n$ directly for the event under study due to lack of observational data. Nevertheless, it is known that in general $n \approx \left(0.1 -1.0\right) \times 10^{10}$~cm$^{-3}$ in electron acceleration sites  \cite{Aschwanden97}. Thus, we conclude that in our case $n \simeq E^{2}/KL$. From Equation~(\ref{eq1}) this means that non-thermal electrons, at least with energy close to the possible low-energy cutoff $E_{\rm{c}}$, could lose a significant fraction of their initial energy in the course of precipitation from the acceleration site to the chromosphere.

This also means that we cannot expand the right-hand side of Equation~(\ref{eq1}) in a series of powers $\left(KnLE^{-2}\right)$. To estimate $\alpha = I_{1}/I_{0}$ we have to find the integral $I_{1}$ of the binomial differentials 
\begin{equation}
 I_{1} \approx A \int^{+\infty}_{E_{\rm{c}}} E^{1-\delta} \left[1 - \left(1-2KnLE^{-2}\right)^{1/2} \right] dE .
\label{eq2}
\end{equation}
This can be solved analytically only if $\left(\delta/2 -1 \right) \in \mathbb{Z}$ or $\left(\delta/2 -1/2\right) \in \mathbb{Z}$ \cite{Gradshteyn07}. We restrict our further calculations to the particular case of $\delta = 4$, which is a typical value for solar flares. In this case we find
\begin{equation}
 I_{1} \approx I_{0} - \frac{A}{6KnL} \left[1-\left(1-2KnLE_{\rm{c}}^{-2}\right)^{3/2}\right],
\label{eq3}
\end{equation}
and consequently
\begin{equation}
 \alpha \approx 1 - \frac{E_{\rm{c}}^{2}}{3KnL} \left[1-\left(1-2KnLE_{\rm{c}}^{-2}\right)^{3/2}\right] .
\label{eq4}
\end{equation}

\begin{figure}
\centerline{\includegraphics[width=0.98\textwidth, bb=90 370 495 786, clip=]{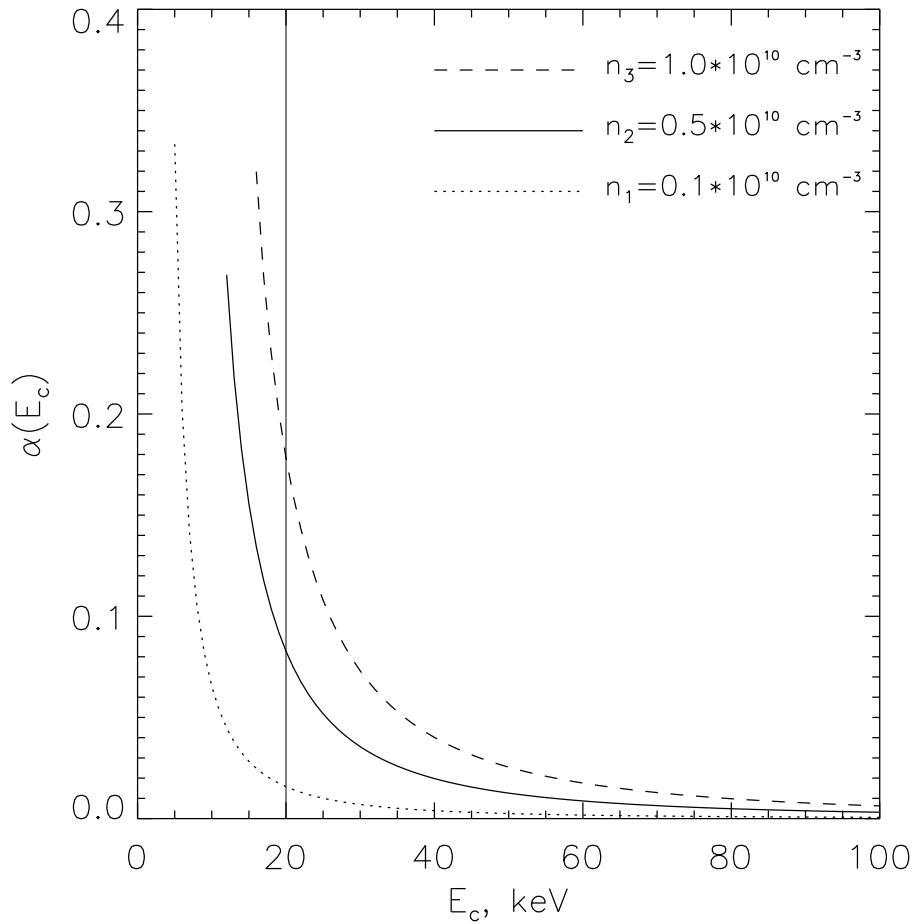}
            }
\caption{Fraction $\alpha$ of the total initial energy of non-thermal electrons lost in the coronal part of the flare loops due to Coulomb collisions with ambient plasma as a function of electron's low-energy cutoff $E_{\rm{c}}$. The dotted, solid, and dashed lines correspond to plasma density of $n_{1}=0.1\times10^{10}$~cm$^{-3}$, $n_{2}=0.5\times10^{10}$~cm$^{-3}$, and $n_{3}=1.0\times10^{10}$~cm$^{-3}$, respectively. }
\label{fig4}
\end{figure}

Figure~\ref{fig4} shows $\alpha\left(E_{\rm{c}} \right)$ curves for the middle $n_{2}=0.5\times10^{10}$~cm$^{-3}$ and two uttermost reasonable values $n_{1}=0.1\times10^{10}$~cm$^{-3}$ and $n_{3}=1.0\times10^{10}$~cm$^{-3}$ of plasma density in the acceleration site \cite{Aschwanden97}, and for the middle value of the possible flare loop semi-length $L_{\rm{loop}}/2\approx L \approx 4\times 10^{9}$~cm. It is seen that a population of nonthermal electrons with low-energy cutoff $E_{\rm{c}} \gtrsim 20$~keV loses only a small fraction of its total initial energy in coronal levels. The situation is opposite when $E_{\rm{c}} \lesssim 10$~keV. For $10 \lesssim E_{\rm{c}} \lesssim 20$~keV we have an intermediate situation --- accelerated electrons lose comparable amount of their total energy both in the coronal and chromospheric parts of flare loops. We will discuss only the first limiting situation when $E_{\rm{c}} \gtrsim 20$~keV. The second limiting situation, when $E_{\rm{c}} \lesssim 10$~keV, contradicts the observations (Figure~\ref{fig1}), since in this case we would expect the same or even greater emission measure during the ``delayed'' bursts with respect to the flare impulsive phase.  
 
We will assume that non-thermal electrons lose the bulk of their energy at the footpoints of the flare loops due to Coulomb collisions with its plasma, and, thus, the majority of the observed soft X-ray emission is radiated by evaporating chromospheric plasma. Energy losses of sub-relativistic non-thermal electrons due to bremsstrahlung can be neglected in comparison with energy losses due to collisions, since 
\begin{equation}
\frac{dE_{\rm{rad}}}{dE_{\rm{col}}} \simeq \frac{1}{\pi} \frac{e^{2}}{\hbar c} \left(\frac{v}{c}\right)^2 \frac{1}{\lambda_{\rm{ee}}} \ll 1   
\label{eqeratio}
\end{equation}
(\eg{}, \opencite{Jackson62}).
  
Let us now estimate the relationship between the soft X-ray emission radiated by evaporated plasma and the flare loop length using the following further assumptions:     
\begin{enumerate}
\item{Soft X-ray emission from the flare loop is thermal bremsstrahlung due to the evaporated chromospheric plasma. }
\item{Chromospheric plasma is evaporated due to sharp heating by non-thermal electrons injected in its volume. }
\item{Chromospheric evaporation is a polytropic process $p V^{\gamma}=\rm{const}$, where $p$ is the pressure of evaporated plasma, $V$ is its volume, and $\gamma$ is the polytropic constant, which is not equal to 0, 1, and $\infty$ in our case, \ie{}, the evaporation process is not an isobaric, isothermal, nor isohoric process. }
\item{The evaporated plasma is an ideal gas, which is described by the ideal gas law $p = n k_{\rm{B}} T$, where $n$ is plasma density, $T$ is its temperature, and $k_{\rm{B}}$ is Boltzmann's constant. }
\item{Pressure $p$, temperature $T$, and density $n$ of the evaporated plasma are constant along the flare loop length and radius. }
\item{A flare loop has constant cross-section along its length. }
\item{Density $n_{0}$ of the plasma is constant along the chromospheric part of the flare loop.}
\end{enumerate}

\begin{figure}
\centerline{\includegraphics[width=0.98\textwidth, clip=]{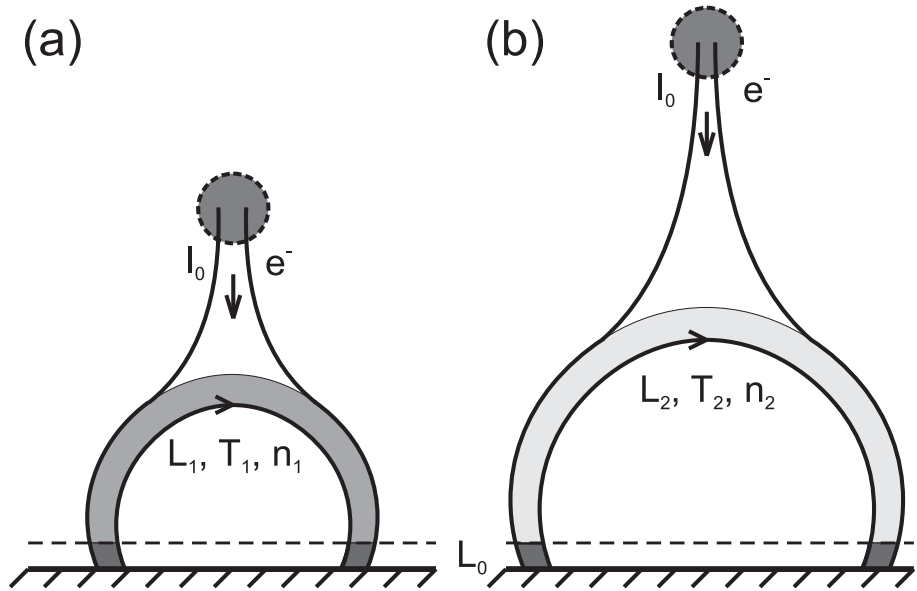}
            }
\caption{Schematic illustration of the discussed interpretation of the observed phenomenon (a) for the impulsive phase of the 26 October 2003 flare and (b) for its ``burst-on-tail''. }
\label{fig5}
\end{figure}

Consider two flare loops with equal cross-sections, but different lengths $L_{1}$ and $L_{2}$ ($L_{1} < L_{2}$). The first case corresponds to the impulsive phase of the event and the second one to the ``delayed'' bursts (Figure~\ref{fig5}). For simplicity we will consider a case when the same population of non-thermal electrons is injected in both loops, thus the same amount of energy is pumped into their plasma. From Equation~(\ref{eq1}) it is evident that in the dense chromosphere the bulk of non-thermal electrons stop traversing over a relatively small length $L_{0} \ll L_{\rm{loop}}/2$. We will assume that non-thermal electrons transmit evenly all their energy to the ambient plasma along this column length $L_{0}$ and the heated chromospheric plasma all evaporates polytropically to the coronal part of the flare loops and emits, in particular, in the soft X-ray range. Let us assume that these processes are instantaneous. In this case, it is possible to estimate analytically relations between the final plasma temperatures $T_{1}/T_{2}$, densities $n_{1}/n_{2}$, emission measure EM$_{1}/$EM$_{2}$, and soft X-ray fluxes $f_{1}\left(E_{1} - E_{2} \right)/f_{2}\left(E_{1} - E_{2} \right)$ integrated over the energy range $\left[E_{1}, E_{2}\right]$, in two different cases of the flare loops with $L_{1}$ and $L_{2}$, as:
\begin{equation}         
 \frac{T_{1}}{T_{2}} \approx \left[\frac{L_{2}}{L_{1}}\right]^{\gamma-1},
\label{eq5}
\end{equation}
\begin{equation}         
 \frac{n_{1}}{n_{2}} \approx \frac{L_{2}}{L_{1}} ,
\label{eq6}
\end{equation}
\begin{equation}         
 \frac{\rm{EM}_{1}}{\rm{EM}_{2}} \approx \frac{L_{2}}{L_{1}} ,
\label{eq7}
\end{equation}
\begin{equation}         
 \frac{f_{1}}{f_{2}} \approx \left[\frac{\exp\left(-E_{1}/k_{\rm{B}}T_{1}\right)-\exp\left(-E_{2}/k_{\rm{B}}T_{1}\right)}{\exp\left(-E_{1}/k_{\rm{B}}T_{2}\right)-\exp\left(-E_{2}/k_{\rm{B}}T_{2}\right)}\right] \left[\frac{L_{2}}{L_{1}}\right]^{\left(5-3\gamma \right)/2} . 
\label{eq8}
\end{equation}

In the event considered $L_{2}/L_{1} \approx 1.7 - 4.5$. It is difficult to estimate this relation more precisely. If we assume an adiabatic process (for simplicity), \ie{}, $\gamma=5/3$, and $T_{1} \approx 19.5$~MK (see Figure~\ref{fig1}), then we find from Equations (\ref{eq5}), (\ref{eq6}), and (\ref{eq7}): $T_{1}/T_{2} \approx 1.4 - 2.7$, $n_{1}/n_{2} \approx 1.7 - 4.5$, EM$_{1}/$EM$_{2} \approx 1.7 - 4.5$. For two standard energy channels of X-ray detectors onboard GOES $\left[E_{1,1}, E_{2,1}\right]$ and $\left[E_{1,2}, E_{2,2}\right]$, where $E_{1,1}=1.55$~keV (8.0 \r{A}) and $E_{2,1}=12.4$~keV (1.0 \r{A}), $E_{1,2}=3.1$~keV (4.0 \r{A}) and $E_{2,2}=24.8$~keV (0.5 \r{A}), we find from Equation (\ref{eq8}): $f_{1,1}/f_{1,2} \approx 1.5 - 4.9$ and $f_{1,2}/f_{2,2} \approx 2.2 - 24.1$, respectively. All these relations are consistent with plasma temperature and emission measure derived from SXR fluxes measured by GOES in the impulsive phase and during the ``burst-on-tail'' (see Figure~\ref{fig1}). This gives good arguments supporting the proposed interpretation. In real flaring regions the process of chromospheric evaporation due to heating by non-thermal electrons is much more complicated (\eg, \opencite{Fisher85a}, \citeyear{Fisher85b}; \opencite{Fisher87}). Nevertheless, we believe that our approach is correct ideologically and might be considered as the first iteration to the problem. 

The estimations presented above show that under special conditions in some flaring regions of LDEs, accelerated electrons can produce intense bursts of non-thermal electromagnetic emission which is not accompanied by significant plasma heating and large fluxes of thermal soft X-ray emission. Recently it was shown by \inlinecite{Fleishman11} that this situation can occur even in impulsive solar flares, which they called ``cold flares''.     

It is possible that both discussed mechanisms are simultaneously responsible for the observed phenomenon. We do not exclude that, along with the increase of spatial scale of the flare loops, the spectrum of non-thermal electrons also became harder during the ``delayed'' bursts and the ``burst-on-tail'', or that the low-energy cutoff of non-thermal electrons increased significantly. Future observations of similar LDEs with pronounced ``bursts-on-tail'' with high spatial and spectral resolution are required for further investigation of the discussed phenomenon. It is also clear that detailed numerical simulations of flare plasma response to accelerated electrons are necessary.   

\subsection{Long Time ``Gap'' of Non-Thermal Emission} 
\label{SS-Gap}
 
There is another interesting aspect of the studied LDE, which as yet remains undiscussed. The gap between the end of the impulsive phase bursts and the first ``delayed'' burst is about 40~min (Figure~\ref{fig1}). This means that electron acceleration to high energies might be interrupted for some reason for about 40~min, and then restored again for another 40~min.  

It seems that the energy release process itself did not fully stop during this gap in the non-thermal electromagnetic emission, since the growth of SXR fluxes, as well as multiple peaks of its time derivative, were well observed during that time. The presence of ongoing flare activity during the gap was also confirmed by the TRACE observations of the growing arcade of magnetic loops. Most likely, the electron accelerator switched to the ``soft'' regime, producing mainly non-thermal electrons of relatively low energies ($E\lesssim 50$~keV), which could effectively heat the active region plasmas emitted in the SXR range, but could not produce HXR ($h \nu \gtrsim 100$~keV) and microwave emissions. Alternatively the accelerator could even shut down during the gap, and the plasma was heated by some other mechanisms, but not by non-thermal electrons. 

Both from a theoretical point of view (see, \eg{}, \inlinecite{Aschwanden02} and \inlinecite{Zharkova11} for a review) and from observations (\eg, \opencite{Sui03}; \opencite{Pick05}; \opencite{Liu08}) it is known that magnetic reconnection in quasi-vertical current sheets, which are expected to be formed during solar eruptions, can lead to acceleration of charged particles. Energy release and charged particle acceleration in real flares are essentially non-stationary. This is clearly confirmed by bursty light curves of flaring non-thermal electromagnetic emissions. Non-stationarity of energy release and acceleration of charged particles in current sheets of eruptive events have, most probably, several physical reasons. Among them are (1) high inhomogeneity of the solar corona and, consequently, high inhomogeneity of current sheets formed during eruptions, (2) intrinsically dynamic features of current sheet evolution, also called as impulsive bursty (\eg, \opencite{Kliem00}) or fractal \cite{Shibata01} reconnection. It is more likely that the gap observed for the 26 October 2003 event was due to the first reason. It seems difficult to explain the long duration of the gap (about 40~min) via bursty reconnection, since this mechanism has a characteristic time scale ranging from fractions of a second to several seconds for typical coronal conditions. In particular, \inlinecite{Warmuth09} have argued that during the ``delayed'' bursts (in our case before them) both the topology and strength of magnetic field in the source of energy release and electron acceleration could change significantly. However, it is not possible to say unambiguously which specific conditions in the flaring region became unfavorable for efficient electron acceleration during the gap in the event studied here. This issue requires further analysis.   

\section{Conclusions}

A prominent example of a long-duration solar event (26 October 2003) with a pronounced delayed sequence of bursts of non-thermal hard X-ray ($h\nu\gtrsim100$~keV) and microwave emission is presented. These bursts were delayed about 40~min with respect to the main impulsive flare phase, and were associated with no appreciable response in the soft X-ray range. This sequence of ``delayed'' bursts itself lasted for more than 40~min, culminating in the most intensive burst of the entire event, which we called the ``burst-on-tail''. We find that the ``delayed'' bursts were generated in the same flaring region as the impulsive bursts, but in magnetic loops a factor of 2-3 larger. 

Based on these observations, an interpretation of the phenomenon is proposed. It is shown that plasma of larger loops, which were involved in the process of energy release during the ``burst-on-tail'', could be heated by a similar population of non-thermal electrons to a lower temperature. This could have resulted in lower values of the emission measure and, consequently, less soft X-ray emission compared with the plasma associated with the relatively small flare loops in the impulsive phase. The estimations made are in good agreement with the observations. 

However, the nature of the long ($\approx40$~min) gap in non-thermal hard X-ray emission between the flare impulsive phase and the sequence of ``delayed'' bursts is not fully understood. Most likely, some physical conditions in the inhomogeneous flaring region became unfavorable for efficient electron acceleration to high energies ($E\gtrsim100$~keV) during that time. However, it is not possible to establish them unambiguously for the event studied. Further investigation of this issue is required.     
  
\begin{acks}
The authors are grateful to the developers of INTEGRAL, GOES, RSTN, NoRP, TRACE and SOHO, which data are used in the paper. They also very thank Dr Andrew Inglis for help in English editing and the anonymous referee for useful comments and remarks. The work was partially supported by the Russian Foundation for Basic Research (project 10-02-01285-a), by the State Contract No.~14.740.11.0086, by the grant NSch-3200.2010.2, and by the Program-22 of the Presidium of the Russian Academy of Sciences.
\end{acks}

  
\bibliographystyle{spr-mp-sola}

\bibliography{references}  

\IfFileExists{\jobname.bbl}{} {\typeout{}
\typeout{****************************************************}
\typeout{****************************************************}
\typeout{** Please run "bibtex \jobname" to obtain} \typeout{**
the bibliography and then re-run LaTeX} \typeout{** twice to fix
the references !}
\typeout{****************************************************}
\typeout{****************************************************}
\typeout{}}

\end{article} 

\end{document}